\documentstyle[12pt,epsf]{article}
\if@twoside
 \else
 \fi
 
 \parskip \medskipamount
\newcommand{\be}{\begin{eqnarray}}
\newcommand{\ee}{\end{eqnarray}}

\begin{document}

\hbox{} \nopagebreak
\vspace{-3cm} \addtolength{\baselineskip}{.8mm} \baselineskip=24pt 
\begin{flushright}
{\sc OUTP-01-02P} \\
hep-ph/0101040
\end{flushright}

\vspace{40pt}
\begin{center}
{\large {\sc {\bf The chicken or the egg; or}}}\\
\vspace{10pt}
{\large {\sc {\bf Who ordered the chiral phase transition?}}} \baselineskip=12pt \vspace{34pt}

Ian I. Kogan $^1$, Alex Kovner$^2$ and Bayram Tekin $^1$ \vspace{24pt}

$^1$Theoretical Physics, Oxford University,\\ 
1 Keble Road, Oxford, OX1 3NP,
UK\\
\vspace{10pt}
$^2$ Department of Mathematics and Statistics,\\ 
University of Plymouth,
Plymouth PL4 8AA, UK\\

\vspace{40pt}
\end{center}

\vspace{20pt}
\begin{abstract}
We draw an analogy between the deconfining transition in the 2+1 
dimensional Georgi-Glashow model and the chiral phase transition in
3+1 dimensional QCD. Based on the detailed analysis of
the former \cite{dkkt} 
we suggest that the chiral symmetry restoration in QCD at high temperature
is driven by the thermal ensemble of baryons and antibaryons. The chiral symmetry
is restored when roughly half of the volume is occupied by the baryons.
Surprisingly enough, even though baryons are rather heavy, a crude estimate
for the critical temperature gives $T_c=180$ Mev. In this scenario the 
binding of the instantons is not the cause but rather a consequence of the 
chiral symmetry restoration.
\end{abstract}

\vfill

\newpage

\section{Introduction.}
In this paper we  suggest that the chiral symmetry restoration
in QCD at 
high 
temperature is driven by the presence of baryons in the thermal ensemble.
In this scenario the chiral symmetry is restored at the temperature
at which the density of the baryons (and antibaryons) in the thermal 
ensemble is 
large enough so
that they start to overlap in space.

There are two main properties of the baryon that render this proposal
physically sensible. First, chiral properties of
the baryon are the same as of a skyrmion in the effective chiral Lagrangian.
That is, inside the baryon the chiral condensate has the 
opposite sign to that in the vacuum\cite{skyrme}. 
Thus if half of the space is filled with baryons, the average value of the 
chiral condensate
vanishes and the chiral symmetry is restored.
The second crucial property is that even though the baryons are heavy,
they 
are
spatially very large. Thus the temperature at which the baryons start 
overlapping
in space is not of the order of their mass, but is significantly smaller.
We will present some rough estimates of this temperature later on and will 
show that it is in the ball-park of 180 Mev.

This mechanism is in a way a competing mechanism to the instanton
binding, 
which 
has been advocated and studied in \cite{shuryak}. According to the instanton 
binding scenario, it is the binding of instantons into ``molecules"
that 
drives the
restoration of the chiral symmetry. In our scenario the symmetry is restored
practically independently of the instanton dynamics. However once the 
symmetry restoration has taken place, the instantons are indeed bound in pairs
by linear ``potential". Thus the instanton binding is not the cause,
but rather the consequence of the chiral symmetry restoration.

Before discussing QCD we would like to make our point on a simpler example,
where one can show analytically that a similar mechanism is indeed responsible
for a thermal phase transition. The case in point is the Georgi-Glashow model
in 2+1 dimensions, and the transition is the deconfining phase transition.
Many years ago Polyakov \cite{polyakov} showed that this theory is
confining. Ever since this model has been used as a test ground for 
various ideas about the dynamics of confinement in 3+1 dimensional theories.
It may perhaps seem surprising that we will be using 
it as a prototypical example
for chiral rather than confining dynamics. But then again this remarkable model
is full of surprises!

Let us first explain in what sense the dynamics of the 3D
Georgi-Glashow model is similar to the chiral dynamics of QCD.

\section{The Georgi-Glashow model - symmetries, anomalies, instantons 
and ``baryons".}

Consider the $SU(2)$ gauge theory with a scalar field in the
adjoint representation in 2+1 dimensions.
\be
S= -{1\over 2g^2}\int d^3x \mbox{tr}\left(F_{\mu \nu}F^{\mu 
\nu}\right) 
+ \int d^3x \left[{1\over 2}(D_\mu h^a)^2 +{\lambda
\over  4}(h^a h^a - v^2)^2 \right] 
\label{model1}
\ee
Here  $A_\mu = {i\over 2} A^a_\mu \tau^a$, 
$F_{\mu\nu} = \partial_\mu A_\nu  -\partial_\nu A_\mu +[A_\mu, A_\nu]$,
$h = {i\over 2} h^a \tau^a$, and $D_\mu h = \partial_\mu h + [A_\mu, h]$.

In the weakly coupled
regime $v\gg g^2$, perturbatively the gauge group is
broken to $U(1)$ by the large expectation value of the Higgs field.
The photon associated with the unbroken subgroup is massless whereas
the Higgs and the other two gauge bosons $W^\pm$ are heavy with the masses
\begin{equation}
M^2_H= 2\lambda v^2, \hskip 1.5 cm M^2_W=g^2v^2.
\end{equation}
Thus perturbatively the theory behaves very much like
electrodynamics with spin one charged matter.

This theory has a global symmetry which will play a very prominent
role in the following discussion. This is the magnetic symmetry
\cite{thooft, kovner}.
Classically the following gauge invariant current is conserved
\begin{equation}
\tilde{F}^{\mu }=\epsilon ^{\mu \nu \lambda}\hat{h}_{a} F^a_{\nu \lambda} -\frac{1}{e}\epsilon ^{\mu \nu \lambda
}\epsilon ^{abc}\hat{h}_{a}({\cal D}_{\nu }\hat{h})^{b}({\cal D}_{\lambda }
\hat{h})^{c}  \label{F}
\end{equation}
where $\hat h^a=h^a/|h|$.
This current defines a conserved charge through $\Phi=\int d^2x \tilde
F_0(x)$. The
continuous $U_M(1)$ magnetic symmetry generated by this charge 
is spontaneously broken in the vacuum, and the
massless photon is the Goldstone boson which reflects this breaking in
the spectrum.

However there are important quantum nonperturbative effects that
change this picture in significant ways.
Those are of course the effects of monopole-instantons. The theory
supports stable Euclidean configurations with finite action
\be
&&h^a(\vec{x})=\hat x^a h(r) \cr
&&A^a_\mu(\vec{x})= {1\over r} \left[ \epsilon_{a\mu
\nu}\hat{x}^\nu(1-\phi_1)+ \delta^{a\mu}\phi_2 +(r A-\phi_2)\hat{x}^a 
\hat{x}_\mu \right]       
\label{configuration1}       
\ee 
where $\hat x^a = x^a/r$. 
In the presence of such a monopole the magnetic current is not
conserved, but rather has a non-vanishing divergence proportional to
the monopole density.
\be
\partial_\mu\tilde F_\mu={4\pi\over g}\rho
\ee
The $U_M(1)$ magnetic symmetry is thus {\it anomalous} in the quantum
theory.
It can be shown \cite{kovner} 
that only the discrete $Z_2$ subgroup is unaffected by
anomaly and thus remains a symmetry in the full quantum theory.

Due to this anomaly the photon becomes a pseudo-Goldstone boson and acquires
a finite mass. This mass is proportional to the density of monopoles,
and is exponentially small at weak coupling,
$m_{ph}^2\propto\exp\{-4\pi M_W/g^2\}$.

Another effect of the monopoles is confinement of $W^\pm$ bosons.
The physically transparent way to see this is to consider the
effective low energy description of the model.
As discussed in detail in \cite{kovner,dkkt} 
the relevant degree of freedom at low energies is the scalar field $V$
that creates a magnetic
vortex of flux $2\pi/g$. 
Under the anomalous magnetic rotation by the angle $\alpha$ it
transforms as
\be
V\rightarrow e^{i{2\pi\over g}\alpha}V
\ee
so that the conserved $Z_2$ subgroup ($\alpha=g/2$) 
acts on it by the sign change.
The low energy effective
Lagrangian in terms of the vortex field is
\begin{equation}
{\cal L}=\partial_\mu V^*\partial^\mu V -\lambda(V^*V-\mu^2)^2 -{\frac{m^2}{4%
}}(V^2 +V^{*2}) + \zeta(\epsilon_{\mu\nu\lambda}\partial_\nu V^*
\partial_\lambda V)^2  \label{ldualgg}
\end{equation}
The coupling constants in eq.(\ref{ldualgg}) are determined in the weakly
coupled region from perturbation theory and dilute monopole gas
approximation. In the weakly coupled region (assuming that the $W^\pm$
bosons are much lighter than the Higgs particle) we have 
\begin{eqnarray}  \label{couplings}
&&\mu^2={\frac{g^2}{8\pi^2}} \hskip 1 cm \lambda={\frac{2\pi^2 M^2_W}{g^2}} \\
&&m=m_{ph} \hskip 1.3 cm \zeta\propto {\frac{1}{g^4M_W}}  \nonumber
\end{eqnarray}
Here $m_{ph}$ is the exponentially small nonperturbative photon
mass calculated by Polyakov \cite{polyakov}.

As discussed extensively in \cite{kovner} the $W$-bosons appear in
this low energy description as solitons. They carry a unit winding
number of the field $V$. Placing $W$ at a point $x$ forces the phase
of $V$ to wind along any curve that surrounds $x$. Due to the fact
that the global symmetry of the effective Lagrangian is $Z_2$ and not
$U(1)$, the lowest energy configuration that carries a unit winding is
not rotationally symmetric hedgehog, but rather a quasi one dimensional
string-like configuration see Fig.1. 
\begin{figure}
%[htb]
\begin{center}
\epsfxsize=3in
\epsfbox{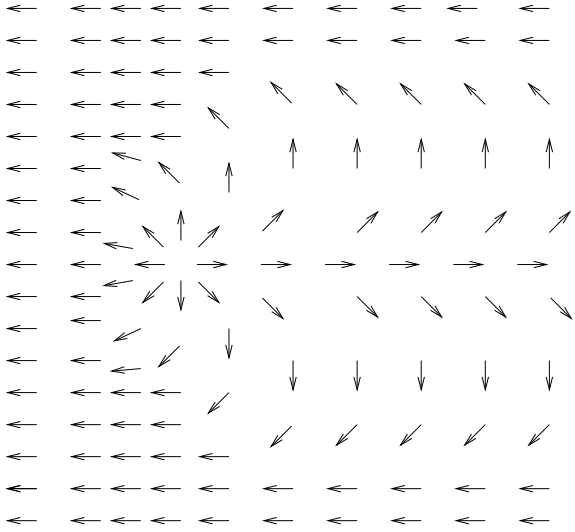}
\caption{The  string like configuration of the field $V$
in the state of unit charge ($W$-boson).}
\end{center}
\end{figure}

The energy of this configuration
is proportional to the length of the string with the string tension
parametrically of order
$g^2m_{ph}$.
A pair of heavy $W^+$ and $W^-$ separated by a distance
$R>1/m_{ph}$ is connected by a string and is
confined.
In fact a more careful analysis \cite{kovner1} reveals that when the distance
$R$ is large this ``adjoint''  string splits in two
``fundamental'' ones. The fundamental string in the effective
Lagrangian appears as a domain wall separating two possible vacuum
states of the field $V$, which are 
degenerate due to spontaneous breaking of the
magnetic $Z_2$. As shown in \cite{kovner1} these fundamental strings
repel each other, and thus it is energetically favorable for the
adjoint string to split into two fundamental ones. 
Due to the linear confinement, the $W$ bosons do not appear in the
spectrum. The actual finite energy excitations are heavy $W^+$-$W^-$
bound states. Such a state naturally looks 
like a domain of one vacuum inside the
other one see Fig.2.\footnote{The domain walls themselves of course have a
finite thickness of order of the inverse photon mass.}
Thus inside the bound state the value of the order parameter $V$ has
the opposite sign that in the surrounding vacuum.
\begin{figure}
%[htb]
\begin{center}
\epsfxsize=3in
\epsfbox{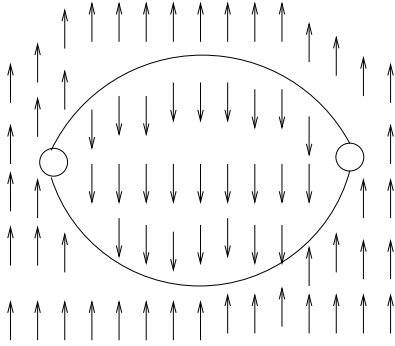}
\caption{The  $W^+$-$W^-$ bound state as the domain of the other
vacuum.}
\end{center}
\end{figure}

Many elements in the structure just discussed are very similar to QCD
with massless fermions. The analogy we have in mind is the following.

$\bullet$ Classical axial $U_A(1)$ symmetry $\leftrightarrow$ Classical magnetic
$U_M(1)$ symmetry .

$\bullet$ Axial anomaly due to instantons $\leftrightarrow$ Magnetic anomaly due
to monopoles.

$\bullet$ Non-anomalous $Z_{N_f}$ subgroup of $U_A(1)$ 
$\leftrightarrow$ Non-anomalous $Z_{2}$ subgroup of $U_M(1)$.

$\bullet$ Spontaneous breaking of $Z_{N_f}$ by the chiral condensate
$<\bar\psi\psi>$$\leftrightarrow$ Spontaneous breaking of $Z_2$ by
the vortex condensate $<V>$.

$\bullet$ Heavy baryons-skyrmions: pockets of the other $Z_{N_f}$ vacuum 
$\leftrightarrow$ Heavy $W^\pm$ bound states: pockets of the other
$Z_2$ vacuum.

There is another important similarity between the baryons and the
bound states in the Georgi - Glashow model. Both are heavy, but
spatially large. In the Georgi - Glashow model, the mass of the bound
state is roughly $M=2M_W$, while the size $D$ is of the order of the inverse
photon mass. Thus there exists a parametric inequality
$M>>D^{-1}$. In QCD of course there is no parametric inequality of
this type, since the theory does not have a dimensionless coupling
constant. Nevertheless the mass of the nucleon (940 Mev) is about ten
times bigger than its inverse diameter (the radius is R=.88 fm)\cite{radius}.

\section{The deconfining phase transition.}

While the zero temperature properties of the Georgi - Glashow model just described
have been known for quite a while, the finite temperature deconfining
phase transition has been studied only very recently\cite{dkkt}.
The dynamics of this transition is quite interesting and turned out 
to be somewhat unexpected.

A natural, but as it turns out misleading way to think about the deconfining
transition is in terms of the dynamics of the monopole ``plasma". At zero
temperature the potential between monopoles is the 3D Coulomb potential
$1/r$ and therefore the monopole gas is in the ``plasma" phase. 
At finite temperature, when one of the dimensions is compactified the
potential at distances $r>T$ turns into two dimensional Coulomb, that 
is logarithmic. The strength of the logarithmic interaction is proportional to 
the temperature, and at temperature $T_{BKT}={g^2/2\pi}$
the monopoles bind in pairs via the Berezinsky-Kosterlitz-Thouless mechansim. 
Above this temperature the monopole gas is 
in the molecular phase. Since at zero temperature it is the monopole plasma 
effects that are responsible for confinement, one may be tempted to 
conclude that this BKT transition in the monopole gas is indeed the
deconfining transition of the Georgi-Glashow model \cite{az}.

A more careful analysis however shows that the situation is much more interesting.
The dynamics of the transition is completely different, and the critical 
temperature is half the value predicted by the monopole binding 
mechanism\cite{dkkt}.
The real culprit are not the monopoles but rather the $W^\pm$ bosons, or 
equivalently their bound states. 
It may seem at first that $W$ can not possibly affect the transition, since
they are extremely heavy.
However, even though their fugacity is very small
at all temperatures of interest ($\exp\{-M_W/T\}$ with $T\propto g^2$), their
effect is long range and therefore strongly affects the infrared properties 
of the system. As should be clear from the preceding discussion, 
the presence of $W$ tends to disorder 
the vortex field $V$, since inside the confining
strings which are attached to $W$ the phase of $V$ has maximal possible variations. 
Thus when the
density of $W$'s is large enough, the vacuum of $V$ becomes disordered and the 
magnetic $Z_2$ symmetry restoration occurs. The magnetic
symmetry restoration is indeed equivalent to deconfinement as discussed in detail
in \cite{kovner2}. The analysis of \cite{dkkt}
shows that the transition occurs at the temperature at which the fugacity
of the $W$ bosons becomes equal to the ``fugacity" of monopoles and in the BPS limit
one has
\be
\exp\{-M_W/T_C\}=\exp\{-4\pi M_W/g^2\}, \ \ \ \ \ \ \ \ T_C={g^2\over 4\pi}
\ee
At this temperature the mean distance between the $W$ bosons in the thermal 
ensemble becomes equal (comparable) to the inverse mass of the photon.
This point has a special significance in terms of the bound states of $W^+$ 
and $W^-$.
As explained above these bounds states are essentially domains of the second
vacuum ($<V>=-\mu$) inside the bulk vacuum $<V>=\mu$. The size of
these domains is of the order of the inverse photon mass. Thus the 
transition occurs 
precisely at the temperature at which a finite fraction of the volume
of the system is occupied by these domains of the second vacuum\footnote{The 
exact fraction
of the volume was not calculated in \cite{dkkt}. It however follows from 
the results of \cite{dkkt} that this fraction is finite and not suppressed by an
exponential factor of the type $\exp\{-AM_W/g^2\}$. Since the dependence of
the $W$ fugacity on the inverse temperature is exponential, this is enough
to determine the critical temperature up to sub-leading corrections in powers of
$g^2/M_W$.}.
Indeed physically this is very reasonable. At the point when 
$<V>=\mu$ in half of the volume 
and $<V>=-\mu$ in the other half , the
expectation value of $V$ over the whole volume, and thus over the thermal ensemble 
vanishes. This is precisely where the symmetry restoring transition has to occur.

It was also shown in \cite{dkkt} that once the transition occurs, the
potential between monopoles changes qualitatively. It becomes linear 
at large distances.
Thus it is indeed true that the monopoles are bound in pairs above the transition.
However this binding does not drive the phase transition but is 
rather the consequence of the transition which is driven by an entirely physically
different mechanism - the overlap of the bound states in the thermal ensemble.

This picture of the transition is very simple and has a certain feel of
universality about it. It seems very likely that a similar mechanism
can operate in other cases. 
In particular in view of the similarities between the Georgi-Glashow model
and chirally invariant QCD, we think that it is very interesting to explore
whether the same mechanism is responsible for the chiral symmetry restoration.
In the next section we will make some very rough estimates of the transition
temperature assuming this is indeed the case.

\section{Baryon driven chiral symmetry restoration.}

Thus the picture of the chiral symmetry restoring phase transition we
advocate is
the following.
At finite temperature the thermal ensemble contains some number of
baryons and antibaryons. Inside the baryon the sign of the chiral
condensate $\bar\psi\psi$ is opposite to that in the vacuum. As
temperature increases the density of the baryons grows. At some point
the
density is large enough so that half of the volume is filled by the
chiral condensate of the opposite sign. At this temperature the order
parameter
averaged over the thermal ensemble vanishes and the chiral symmetry is
restored.

The factor that works against the symmetry restoration is the high
mass of the baryon. On the other hand there are several factors that 
help. First, the size of the baryon is large. In the following
estimates we will use for the radius of the baryon $R=.88$
fm\cite{radius}. Strictly speaking this is
the charge radius, however the radius of the region of the
wrong-sign-condensate is very similar \cite{petrov}. Second, the entropy of the
baryons is quite large. In the two flavor case we will take into
account nucleon and delta, including their spin and isospin degrees of
freedom. Third, the radius of the baryon itself depends on temperature
and is believed to grow as the temperature rises. 
Although no reliable calculation of the swelling of the baryon
size exists, it is reasonable to expect that the size increases by
about 10-20\% at the critical temperature due to the decrease of
$F_\pi$. 
We will try to model this
last effect in a very simplistic way.

To estimate the critical temperature we approximate the baryon
ensemble by a non-relativistic ensemble of free noninteracting
particles.
The density of particles in such an ensemble is given by
\begin{equation}
n(T) = \sum_iN_i( { M_i T\over 2\pi})^{3/2} 
    e^{-{M_i\over T}}
\end{equation}
where $M_i$ is the mass of the particle of species $i$ and $N_i$ is
the number of degrees of freedom with this mass.

We estimate the critical temperature by equating the fraction of the
volume occupied by the particle to 1/2. In all the estimates we take
the radius of all the relevant baryons to be equal.
We will consider in the following the cases of 2 and 3 massless
flavors as well as the realistic case of the massive strange quark.

Let us first consider the two flavor case. 
The only baryons important for the transition are the nucleon and the delta
with  $M_n = 938$ MeV and $M_{\Delta}= 1232$ MeV. We have checked
numerically that including the Roper resonance does not affect the results.
The fraction of the volume occupied by the nucleons and deltas
at temperature $T$
is 
\begin{equation}
f(T) = 8{4\pi R(T)^3\over 3}\,\, ( { M_n T\over 2\pi})^{3/2}\,\, 
    e^{-{M_n\over T}} \left \{ 1 + 4({M_{\Delta}\over M_n})^{3/2} 
e^{{M_n - M_{\Delta}\over T}} \right \} 
\end{equation}
where the entropy factor is $2(2S+1)(2I+1)$ for particle-antiparticle,
spin and isospin degrees of freedom.
In this formula we allowed for the temperature dependence of the
nuclear radius.
Neglecting this effect first, we plot the fraction $f(T)$ in Fig.3.
The striking feature of this plot is that all the action happens
in the relatively narrow window between $T=150$ Mev and $T=215$ Mev. 
Note that this temperature range is indeed much lower than the baryon
mass and is in the right ball-park for the chiral phase transition.
The value of the critical temperature we extract from this graph is
$T_c=213$ Mev.

\begin{figure}
%[htb]
\begin{center}
\epsfxsize=3in
\epsfbox{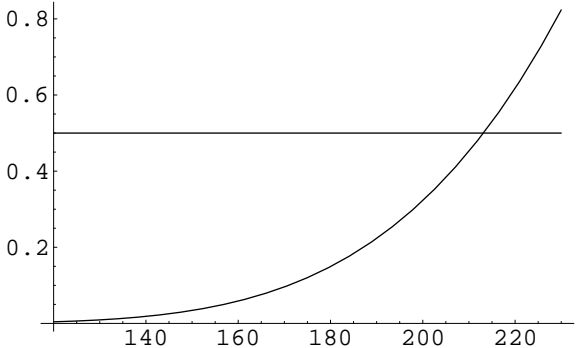}
%\centering{\epsfig{file=deltas.eps}}
\caption{In two flavor QCD the fraction of the volume occupied by the nucleons and
deltas as function of temperature. 
The phase transition temperature is $T= 213$ MeV. The radius of particles is assumed to be
temperature independent.} 
\end{center}
\end{figure}

We next try to take into account the swelling of the baryon radius
with temperature. Our simple ansatz for this dependence is 
\begin{equation}
R(T) = R(0) + {1\over  M_\sigma(T)} - {1\over  M_\sigma(0)}
\label{radius}
\end{equation}
with $M_\sigma$ - the mass of the $\sigma$-particle.
\begin{equation}
M_\sigma(T) = {F_\pi(T)\over F_\pi(0)} M_\sigma(0)
\label{smass}
\end{equation}

The rationale for this is the following. 
The chiral order parameter $\bar\psi\psi$ couples directly to the 
$\sigma$ - particle.
Inside the nucleon the chiral
order parameter has a negative sign. It has to relax to its vacuum
value on the outside. This relaxation happens either through the
``phase rotation'' if $\sigma$-particle is very heavy, or through the
change in the $\sigma$-field itself. In the latter case the distance
over which it happens should be equal to the inverse $\sigma$-mass. Closer to
the phase transition, $\sigma$ becomes light and effective in the
relaxation of the order parameter field. Eq.(\ref{radius}) is a simple
interpolation between the low temperature situation, where $\sigma$
is heavy and unimportant and the closer-to-criticality situation,
where it does indeed contribute significantly to the size. The formula
eq.(\ref{smass}) is just the simple linear $\sigma$ - model type
relation.
We do not insist that eq.(\ref{radius}) has any precision, but
we believe that it gives a rough estimate of the effect\footnote{We
note that a similar effect of the change of $F_\pi$
with temperature and the associate change in the size of the bound
state is also present in the Georgi - Glashow model. Just like in QCD
it is due to thermal fluctuations of the light particles, which
in 3D are light photons. The reason we did
not discuss it here, is that it is parametrically sub-leading. That is, it
affects the correction to the value of the critical temperature at
relative order $g^2/M_W$. Since QCD does not have a free
parameter, the effect is likely to be more important in QCD and
therefore should be taken into account. The effect of pions 
should disappear in $SU(N)$ theories for large
$N$, since at large $N$ both the inetraction of pions is weak and the
ratio of the proton mass to its inverse size is parametrically large. }. 
We take  $M_\sigma(0)= 600$ MeV and
$F_\pi(0) = 93$ MeV.

To use this relation we still need to know the dependence of $F_\pi$
on the temperature.  In the lowest order in temperature it is given by
 \cite{Leutwyler}
\begin{equation}
F_\pi(T)= F_\pi(0) ( 1 - {T^2\over 12  F_\pi(0)^2} )
\label{pert}
\end{equation}
The Pad\'{e} resummed expression which should better represent the
situation closer to criticality ($T_c$) has been proposed in \cite{Kapusta}.
\begin{equation}
{F_\pi^2(T)\over F_\pi^2(0)} = {{1- T^2/T_c^2} \over 1- {2\over 3}(T^2/T_c^2)
( 1- T^2/T_c^2)}
\label{kapusta}
\end{equation}
This formula assumes that the symmetry is $O(4) = SU(2)\times SU(2)$. 
Using eq.(\ref{pert}) the graph for the fraction $f(T)$ is given on
Fig. 4. The critical temperature is $T= 195.5 $ Mev. 

\begin{figure}
%[htb]
\begin{center}
\epsfxsize=3in
\epsfbox{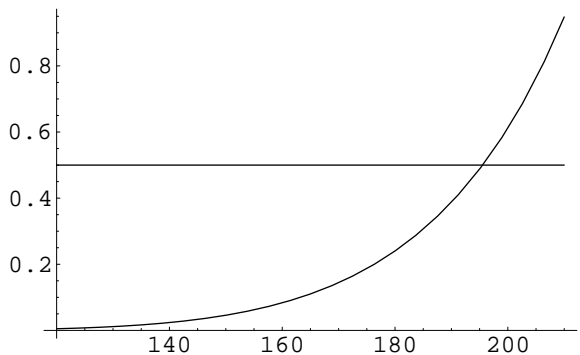}
\caption{ In two flavor QCD the fraction of the volume occupied by the nucleons and
deltas. The dependence of the radius on temperature is given by eq.(\ref{pert}).
The transition temperature is $T= 195.5$ MeV } 
\end{center}
\end{figure}

Using $T_c=195.5$ in
eq.(\ref{kapusta})
we obtain Fig.5 with the critical temperature $T= 179$ Mev. 
Thus the swelling of the baryon radius has an effect of reducing the
critical temperature by about 15\%.
\begin{figure}
%[htb]
\begin{center}
\epsfxsize=3in
\epsfbox{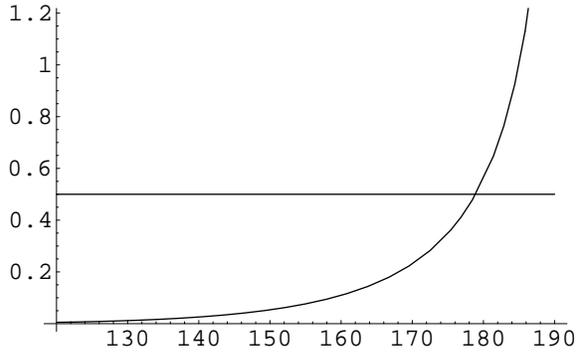}
\caption{ Same as in Fig.4 but with eq.(\ref{kapusta}) and $T_c=195.5$.
The transition temperature is $T= 179$ MeV } 
\end{center}
\end{figure}

It is interesting to see how the value of the critical temperature
depends on the number of flavors. 
For $N_f=3$ case we should consider the baryon octet and decouplet.
To get a rough idea here we will neglect the temperature dependence of
the radius. In the idealized chirally symmetric three flavor case
we take the octet mass as the mass of the nucleon and the decouplet
mass as the mass of the delta. The resulting curve is plotted on
Fig. 6.

\begin{figure}[htb]
\begin{center}
\epsfxsize=3in
\epsfbox{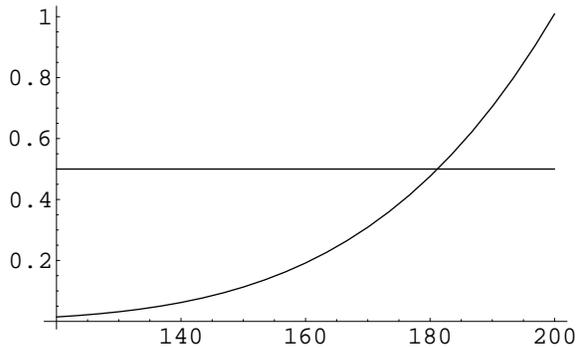}
\caption{ Fraction of the volume occupied by the baryons for three
massless flavours. 
The transition temperature is $T= 181$ MeV } 
\end{center}
\end{figure}

The critical temperature is $T=181$ Mev. This is some $30$ Mev lower
than the corresponding value for the $N_f=2$ case. The same
trend exists in the lattice data \cite{lattice}. In our approach
this is easily understandable: it is the direct consequence of having
roughly three times as many active baryons for $N_f=3$ as for $N_f=2$.

Taking instead the physical masses for the octet and decouplet members
we get Fig. 7 with $T_c=195.5$ Mev.
\begin{figure}[htb]
\begin{center}
\epsfxsize=3in
\epsfbox{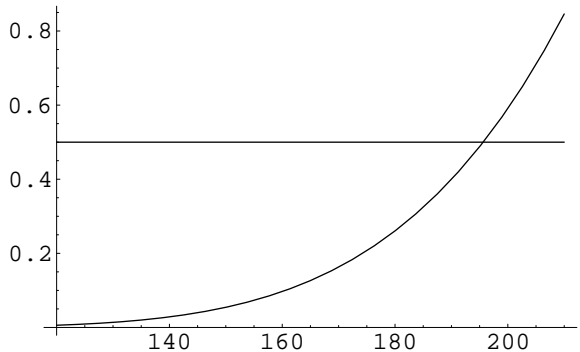}
\caption{ Same as in Fig.6 but with
realistic baryon masses.
The transition temperature is at $T= 195.5$ MeV } 
\end{center}
\end{figure}

\section{Discussion}

The surprising result of our numerical estimates is that 
even though the baryon mass is around 1 Gev,
the baryon overlap
mechanism leads to critical temperature of order $180$
Mev for $N_f=2$, and about $30$ Mev lower for $N_f=3$. These numbers
are perfectly reasonable and are in qualitative agreement with the
lattice results which give $T_c=173\pm 8$ Mev for $N_f=2$ 
and  $T_c=154\pm 8$ Mev for $N_f=3$ \cite{lattice}. Of course our
estimates are very rough and suffer from many uncertainties. For
example, it is not clear that the fraction of the volume must be
really $1/2$. It may be enough to fill a smaller fraction, since the
baryon has a pion tail which itself also contributes to disordering of
the condensate. This would push the value of the critical temperature down.
We also completely neglected the interaction between the baryons,
which start to be important precisely in the region of densities we
are interested in\footnote{We also neglected the fact that baryons are
fermions. This effect is however rather small, and we have checked
numerically that using Fermi-Dirac rather than Boltzmann distribution
changes the value of the critical temperature by about 1 Mev.}.
There is also  an uncertainty of the dependence of the baryon radius
on the temperature. 

Our discussion of effects due to the thermal bath 
of mesons has been very rudimentary.
Partly this effect has been taken into account by 
allowing for the temperature dependence of $f_\pi$(for more details see \cite{Leutwyler},
\cite{Kapusta} and references therein) which leads to the
renormalization of the
 baryon parameters. This reduction in the value of $f_\pi$
is due to direct disordering of the chiral vacuum by the thermal pions.
The fact that the critical temperature we obtain is always lower than
the input $T_c$ in eq.(\ref{kapusta}) is in our view an indication
that the disorder due to baryons takes precedence over the direct pion
effects.
The thermal production of vector and axial mesons we believe is less
relevant since they are almost as heavy as baryons, but contrary to
baryons have no direct disordering effect on the vacuum.

Since at this time we do not know 
how to take these effects into account in a well
defined calculational framework, our discussion 
has been rather qualitative. It is however
encouraging that the numbers fall in the right ball-park\footnote{Due
to the uncertainties in our estimates one has to be careful using them
in some situations. For example a straightforward application of our 
argument would lead one to conclude that in the ordinary
nuclear matter, chiral symmetry should be restored already at zero
temperature, since the
packing fraction of the baryons is close to one. In fact, however the
critical density at which the chiral symmetry is restored is thought
to
be 2.5 to 3 times the nuclear matter density\cite{smilga}. 
There is a significant
difference however between the finite temperature and finite density
situations. At finite temperature, due to the  Boltzmann factor the
dependence of the temperature on the packing fraction is essentialy
logarithmic. Thus a change of order one in the packing fraction does
not lead to significant change in the value of the critical
temperature. On the other hand critical density is directly
proportional to the packing fraction, and is thus very sensitive to
any changes in it. One can certainly imagine dynamical effects which
change the packing fraction from our naive estimates
especially
when a system is relatively dense. For example at finite chemical
potential the size of the region inside the baryon where the order
parameter is negative can shrink.  This is consistent with     
the Skyrme model calculations of the sizes of baryons with higher
baryon number\cite{oka}.
 A change of some 20 percent would be enough to push the
effective packing fraction significantly below one, and thus push
the system deep into chirally symmetric phase.}.

We note that our scenario relates to the instanton binding scenario of
\cite{shuryak} in very much the same way as the actual transition in
3D Georgi-Glashow to the monopole binding scenario \cite{az}. 
In the chirally symmetric phase the potential between 
instantons should be linear whatever the mechanism that drives the
transition is.
This is simple to understand at high temperatures. Consider
the correlation function of some local operator which is not
invariant under the axial $U_A(1)$ but is invariant under the
non-anomalous $Z_{N_f}$ and also under the chiral $SU(N_f)\times
SU(N_f)$.
A good example of such an operator is 't Hooft's effective interaction
vertex \cite{thooft1} $T$.
At high temperature where the instanton gas is dilute and perturbation
theory valid, the calculation of the correlation function
$<T(x)T^*(y)>$ is dominated by the contribution of the instanton-antiinstanton
pair at points $x$ and $y$.
One expects this correlation function to 
approach a constant value at large distance and the leading correction
to be exponential
$<T(x)T^*(y)>\propto [\exp\{-m|x-y|\}+\zeta]$.
In terms of the instanton-antiinstanton
potential
this translates into linear potential which is screened at large
distances.
The screening is the consequence of the ``breaking'' of the string
between instantons, whereby 
an extra instanton-antiinstanton pair appears when the distance $x-y$
is too large\cite{new}.
Thus just like in 3D we expect that the binding of instantons into pairs
in the chirally symmetric phase is a consequence of the phase
transition
even if the transition itself is driven by a noninstanton mechanism.

An interesting property of the mechanism we suggest is a quite
distinct large $N_c$ behavior. The mass of the baryon is proportional
to $N_c$. On the other hand the multiplicity of the lightest baryons 
scales as a power of $N_c$. For example in Skyrme model with 2
flavors one has $I=J = 1/2,3/2,5/2, ..., N_c/2$ baryons with 
masses\footnote{For discussion of more general case including the
strange quark see for
example \cite{manohar}.}
\begin{equation}
M = m_0 N_c + \frac{1}{N_c}m_1 I(I+1)
\end{equation}
The degeneracy
 factor is $(2I+1)^2$ which  (after summation over spins)
leads to the overall extra factor $N_c^3$. 
Thus at large $N_c$ the critical
temperature
predicted by the baryon overlap 
mechanism is $ T_{c} \sim N_c / ln N_c T_0 $ where
$T_0$ is by order of magnitude of $\Lambda_{QCD}$. This 
temperature grows with $N_C$\footnote{Interestingly
although the temperature grows with $N_C$, at large $N_c$ it 
is parametrically smaller than the
baryon mass with the suppression factor $1/ ln N_c$.} On the other hand the
deconfinement phase transition temperature in the pure Yang-Mills
theory
is believed to be $O(1)$ in
the large $N_c$ limit and is parametrically smaller than $T_{c}$.
Thus
it is likely that at some critical number of colors the
chiral symmetry restoration temperature becomes larger than the
deconfinement temperature.

Some arguments have been advanced to the effect that if 
the chiral transition happens at lower temperature, it
also drives deconfinement\cite{satz}.
Thus at small $N_c$ only one transition in QCD with fermions is
observed.
On the other hand if the deconfinement happens earlier, the chiral
symmetry is not necessarily restored above this, first transition. 
In fact the common wisdom is
that the confinement and the chiral symmetry breaking are due to
different sectors of QCD dynamics. If chiral symmetry is still broken
above the deconfining transition, the baryons should still exist there
as bound states of quarks, even though the quarks themselves may be
not confined. Thus the chiral symmetry restoring transition due to the
baryon overlap mechanism can still run out its turn at $T_c=O(N_c)$.
In this case for large enough number of colors the theory will have
two distinct phase transitions: first the deconfining one and later
the chirally restoring one. If the critical $N_c$ is not too large, it
may be possible to see the second transition in lattice simulations.

Another interesting issue is the fate of the hot chirally symmetric
ground state when it is cooled. If the chiral transition 
is second or weakly first order there should be no appreciable
hysteresis and thus during cooling the system should follow through
the same states as during heating but in reverse order. This would
imply production of baryon-antibaryon
pairs in the initial stages of cooling and 
should lead to the production of baryon-rich 
final states in mid rapidity in collision processes which
create quark-gluon plasma in the intermediate stage. 
If the transition is strongly first order there may be large
hysteresis
and cooling could proceed along a different root than heating.

Our suggestion in this paper 
is in large measure motivated by the analogy with the
3D Georgi - Glashow model. We should mention that the analogy is of
course not perfect. The main new element in QCD is the existence of
the continuous chiral symmetry in addition to the non-anomalous
discrete axial one. Thus there are massless pions in the game, which
was not the case in our 3D example. Thus for example the following
question has to be answered. The direct consequence of the baryon
mechanism
is the vanishing of the chiral condensate. It however does not
directly tell us that the pions become massive. In principle the
situation when the order parameter vanishes, but there are still
massless particles around is possible. It is in fact quite generic in
2 dimensional systems due to Coleman theorem. However it seems to us
very unlikely that similar situation 
can be sustained in 4D. Thus we believe that
once the condensate vanishes, pions will acquire a mass.  It
is interesting and 
important to identify a dynamical mechanism through which this is  
achieved\footnote{We are grateful to Victor Petrov for raising this
question and interesting and heated discussions on the subject.}.

Another aspect of QCD dynamics which is different 
compared to 3D Georgi-Glashow theory,
is the role of instantons at zero temperature. In the Georgi-Glashow
model, the monopole-instantons bring about the anomaly in the
magnetic $U_M(1)$ symmetry, but they are not responsible for the
spontaneous breaking of the residual $Z_2$ group. The spontaneous breaking
is there already on the perturbative level. On the other hand in QCD
it is believed that both the anomalous breaking of $U_A(1)$ and the
spontaneous breaking of the residual chiral symmetry are due to
instanton dynamics.
Thus one may be more inclined to believe that the symmetry restoration
transition in QCD is also linked to the instanton physics. However we stress
that it is not at all necessary that the mechanism of the symmetry
restoration is just elimination of the mechanism that
brought about the symmetry breaking in the first place.
Thus although it is logically possible that the instanton binding in
QCD occurs at lower temperature than the baryon overlap, this question
can only be settled by a reliable calculation.
The numerical results of \cite{shuryak} indicate that the critical temperature
for the instanton binding is by about $30$ Mev lower than our
estimate. However given the uncertainties of the calculation of
\cite{shuryak} and even more so the qualitative level of our
estimates here, we feel that much more work has to be done before a
definite conclusion can be drawn on this point.

How does one distinguish between different possible mechanisms is not
an easy question. On the qualitative level however, in the baryon overlap
mechansim the symmetry restoration is due to large fluctuations of the
phase of the order parameter rather than of its magnitude. Thus there
should be a sharp distinction between this scenario and, say the
transition in the linear
$\sigma$ - model. The quantity to measure in this case is the
``square'' of the order parameter, or in the case of two flavours,
rather
the 't Hooft vortex. If the transition is driven by large phase
fluctuations, the average value of the 't Hooft vertex should change
very little across the transition, since it is itself an invariant
operator.
If on the other hand, like in the linear $\sigma$ - model the
magnitude of
the order parameter becomes small at criticality, so should the 't
Hooft vertex. Such a measurement in the lattice gauge theory would be
very interesting\footnote{One has to be careful to appropriately smear
the 't Hooft vertex to get rid of the ultraviolet contributions, which
otherwise can easily blurr the picture}.

We think that the scenario we presented in this note is physically
quite appealing and simple, and thus further work to check its
validity is certainly warranted.

\leftline{\bf Acknowledgments}
We are indebted to Victor Petrov for very interesting and stimulating 
discussions. We also acknowledge discussions with Witek Krasny and
Misha Shifman.
A.K. is supported by PPARC. The research of  I.K. and B. T. are
supported by   PPARC Grant PPA/G/O/1998/00567.

\end{document}